\input harvmac
\writedefs
\sequentialequations

\def\comment#1{}

\def\+{^\dagger}

\def \chi {\chi}

\def\np {{  Nucl. Phys. }}
\def \pl {{  Phys. Lett. }}

\def \prl {{  Phys. Rev. Lett. }}
\def \pr  {{ Phys. Rev. }}


\Title{
 \vbox{\baselineskip10pt
  \hbox{PUPT-1682}
  \hbox{hep-th/9702076}
 }
}
{
 \vbox{
  \centerline{ World Volume Approach to }
  \vskip 0.1 truein
  \centerline{Absorption by Non-dilatonic Branes}
 }
}
\vskip -25 true pt

\centerline{
 Igor R. Klebanov 
 }
\centerline{\it Joseph Henry Laboratories, 
Princeton University, Princeton, NJ  08544}

\bigskip
\bigskip
\centerline {\bf Abstract}
\bigskip
\bigskip
\baselineskip10pt
\noindent
We calculate classical cross sections for absorption of massless
scalars by the extremal 3-branes of type IIB theory, and by 
the extremal 2- and 5-branes of M-theory. The results are compared
with corresponding calculations in the world volume effective
theories. For all three cases we find agreement in the scaling with
the energy and the number of coincident branes. For 3-branes,
whose stringy description is known in detail in terms of
multiple D-branes, the string theoretic
absorption cross section for low energy dilatons is in 
{\it exact} agreement with the classical gravity.
This suggests that scattering from extremal 3-branes is a unitary
process well described by perturbative string theory.

\Date {February 1997}

\noblackbox
\baselineskip 14pt plus 1pt minus 1pt


\lref\mast{J.M.~Maldacena and A.~Strominger, Rutgers preprint RU-96-78, 
hep-th/9609026.}
\lref\mst{J.M.~Maldacena and A.~Strominger, \prl 77 (1996) 428, 
hep-th/9603060.}

\lref\HP{G. Horowitz and J. Polchinski, hep-th/9612146.}

\lref\HK{A. Hashimoto and I.R. Klebanov, 
\pl B381 (1996) 437, hep-th/9604065.}

\lref\JP{J. Polchinski, \prl 75 (1995) 4724, hep-th/9510017.}

\lref\dmII{S. Das and S.D. Mathur,  hep-th/9607149.}

\lref\pope{H. Lu, S. Mukherji, C. Pope and J. Rahmfeld, hep-th/9604127.}

\lref\sv{A.~Strominger and C.~Vafa, \pl B379 (1996) 99, hep-th/9601029.}

\lref\cm{C.G.~Callan and J.M.~Maldacena, \np B472 (1996) 591, hep-th/9602043.}

\lref\ms{J.M.~Maldacena and L.~Susskind, Stanford preprint
SU-ITP-96-12, hep-th/9604042.}

\lref\dmw{A.~Dhar, G.~Mandal and S.~R.~Wadia, 
Tata preprint 
TIFR-TH-96/26, hep-th/9605234.}

\lref\dm{S.R.~Das and S.D.~Mathur,
Nucl. Phys. {\bf B478} (1996) 561, hep-th/9606185. 
 }

\lref\Ark{A. Tseytlin, hep-th/9602064.}

\lref\LS{L. Susskind, hep-th/9309145.}

\lref\hs{G.~Horowitz  and A.~Strominger, Nucl. Phys. B360 (1991) 197.}

\lref\GK{S.S.~Gubser and I.R.~Klebanov, 
Nucl. Phys. B482 (1996) 173, hep-th/9608108.}

\lref\us{I.R.~Klebanov and L.~Thorlacius,
Phys. Lett. B371 (1996) 51, hep-th/9510200;
S.S.~Gubser, A.~Hashimoto, I.R.~Klebanov and J.M.~Maldacena,
\np B472 (1996) 231, hep-th/9601057. }

\lref\gm{R. Garousi and R. Myers, hep-th/9603194.} 

\lref\GKtwo{S.S.~Gubser and I.R.~Klebanov, Phys. Rev. Lett. 77
(1996) 4491, hep-th/9609076. }

\lref\CGKT{
C.G.~Callan, Jr., S.S.~Gubser, I.R.~Klebanov and A.A.~Tseytlin,
hep-th/9610172; 
I.R. Klebanov and M. Krasnitz, hep-th/9612051.}

\lref\dgm{S.~Das, G.~Gibbons and S.~Mathur, hep-th/9609052.}

\lref\GKP{S.S.~Gubser, I.R.~Klebanov and A.W.~Peet, \pr D54 (1996) 3915,
hep-th/9602135; A. Strominger, unpublished notes. }

\lref\KT{I.R. Klebanov and A.A. Tseytlin, \np B475 (1996) 179, 
hep-th/9604166.}

\lref\kt{I.R. Klebanov and A.A. Tseytlin, \np B475 (1996) 165, 
hep-th/9604089.}

\lref\at{ A.A. Tseytlin, \np B475 (1996) 49, hep-th/9604035.}

\lref\CTT{M. Cveti\v c and  A.A.  Tseytlin, 
\pl B366 (1996) 95, hep-th/9510097; \pr D53 (1996) 5619, 
hep-th/9512031.}

\lref\myers{C. Johnson, R. Khuri and R. Myers, 
Phys. Lett. B378 (1996) 78, hep-th/9603061.} 

\lref\bl{V. Balasubramanian and F. Larsen, hep-th/9604189.}

\lref\EW{E. Witten, Nucl. Phys. B460 (1996) 541.}

\lref\guv{R. G\" uven, \pl B276 (1992) 49.}

\lref\ds{M.J. Duff and K.S. Stelle,  Phys. Lett. B253
(1991) 113.}

\lref\Duff{M.J. Duff, H. Lu, C.N. Pope
Phys. Lett. B382 (1996) 73, hep-th/9604052. }

\newsec{Introduction}

Extremal black holes with non-vanishing horizon area may be embedded
into string theory or M-theory using intersecting $p$-branes
\refs{\CTT,\sv,\cm,\mst,\myers,\at,\KT,\bl}. 
These configurations are useful for
a microscopic interpretation of the Bekenstein-Hawking entropy.
Furthermore, a number of emission/absorption calculations agree
with a simple `effective string' model for the dynamics of the
intersection \refs{\cm,\dmw,\dm,\dmII,\GK,\mast,\GKtwo,\CGKT}. 
Unfortunately, at
the moment there is no complete derivation of this model from first
principles. Thus, it is useful to examine simpler configurations
which involve parallel branes only (upon dimensional reduction they
yield black holes with a single type of $U(1)$ charge). 

A microscopic interpretation of
the entropy of near-extremal $p$-branes was first studied in
\refs{\GKP,\kt}. It was found the the scaling of the 
Bekenstein-Hawking entropy with the temperature agrees
with that for a massless gas in $p$ dimensions only for the
`non-dilatonic $p$-branes':
namely, the self-dual 3-brane of the type IIB theory,
and the 2- and 5-branes of M-theory.\foot{Some ideas on how to extend this
agreement to the dilatonic branes \Duff\
were suggested in \pope.}
In \HP\ a way of reconciling the differing scalings 
for the dilatonic branes was proposed.
According to this `correspondence principle' \refs{\LS,\HP}, 
the string theory and the 
semiclassical gravity descriptions are in general
expected to match only at a special value of the temperature, which
corresponds to the horizon curvature 
comparable to the string scale.\foot{
For $N$ parallel D-branes, $N g_{\rm str}$ is of order 1 at the
matching point \HP.}
In \HP\ it was shown that, in all known cases, the stringy and the
Bekenstein-Hawking entropies match at this point up to factors of order
1. Part of the ambiguity in this factor comes from knowing the
matching point only approximately. However, for the non-dilatonic
branes this ambiguity is absent: the matching can be achieved at any
scale because the stringy and the semiclassical entropies have
identical scalings with temperature. This still leaves 
a discrepancy -- the relative factor of 4/3 -- 
for the 3-brane entropy \GKP. In \HP\ a qualitative
explanation of this factor was attributed to strong coupling
effects on the world volume.
In view of the new results that we will
present here, one may wonder if there exists an exact explanation
of the 3-brane entropy in terms of a weakly coupled 
theory (perhaps utilizing the S-duality).

The non-dilatonic branes have a number of special properties.
A notable property of their extremal metrics
is that the transverse part of the 
geometry is non-singular: instead of a singularity we find an infinitely
long throat whose radius is determined by the charge 
(the vanishing of the horizon area is due to the longitudinal
contraction). The metric describing a 
non-dilatonic $p$-brane carrying an 
elementary unit of charge has the spatial curvature bounded from above
by a quantity of order the Planck scale. Thus, for a large number $N$
of coincident branes, the curvature may be made arbitrarily small in
Planck units. For instance, for $N$ D3-branes, the curvature is bounded
by a quantity of order
$$ {1\over \sqrt{N\kappa_{10}}} \sim {1\over \alpha'\sqrt{N g_{\rm str}}}
\ .$$ 
Thus, to suppress the string scale corrections to the classical metric,
we need to take the limit $N g_{\rm str}\rightarrow\infty $.

The tensions of non-dilatonic branes
depend on $g_{\rm str}$ and
$\alpha'$ only through the gravitational constant $\kappa$ in 
the appropriate dimension, which is also the only scale present in the
semiclassical description. Indeed, the D3-brane tension is
$\sim 1/\kappa_{10}$, the M2-brane tension is $\sim 1/\kappa_{11}^{2/3}$,
and the M5-brane tension is $\sim 1/\kappa_{11}^{4/3}$. 
This means that we can compare the expansions of various 
quantities in powers of $\kappa$ between the microscopic and
the semiclassical descriptions. 
It is often said that, in the microscopic description such an expansion
is not tractable because it proceeds in powers of $N g_{\rm str}$,
a quantity that has to be considered very large. We will see, however,
that for the 3-brane
absorption cross section the actual expansion parameter is
\eqn\param{ N\kappa_{10} \omega^4 \sim N g_{\rm str} \alpha'^2 \omega^4
\ ,}
where $\omega$ is the incident energy. Thus, we may consider
a `double scaling limit'
\eqn\dsl{ N g_{\rm str}\rightarrow\infty\ ,
\qquad \omega^2 \alpha'\rightarrow 0\ ,
}
where the expansion parameter \param\ is kept small.
Moreover, the classical absorption cross section is naturally
expanded in powers of $\omega^4\times {\rm curvature}^{-2}$, which
is the same expansion parameter \param\ as the one governing the
string theoretic description of the 3-branes. Thus, the two expansions
of the cross section may indeed be compared, and we will find that
the leading term agrees exactly! In our opinion, this 
provides evidence in favor of
scattering off extremal 3-branes being a unitary process,
well described by perturbative string theory. 

In view of the special properties mentioned above, we believe that
the non-dilatonic branes admit more detailed 
string theory -- semiclassical gravity comparisons than those allowed
in general by the correspondence principle of \refs{\LS,\HP}. 
Also, the world volume dynamics is better understood here than in the
case of intersecting branes, which allows for a calculation from
first principles. Indeed, $N$ parallel D3-branes
are known to be described by a $U(N)$ gauge theory on the world volume \EW.
For multiple M-branes the world volume theory is not known in detail
but, with minimal assumptions about its structure, we will 
be able to make interesting comparisons as well.

The structure of the paper is as follows. In section 2 we calculate 
the classical absorption cross sections for low energy massless
scalars incident at right angles on the non-dilatonic branes.
In section 3 we compare with the cross sections for an incident
scalar to turn into a pair of massless modes on the brane moving
in opposite directions. We find that the scalings with the energy and
the number of branes agree in all cases. For the 3-branes, which is
the only case where we are able to fix the normalizations,
we find exact agreement between the string theoretic and the classical
cross sections. In section 4 we study higher partial waves. We identify
the leading terms in the effective action which convert the incident
scalar into $l+2$ massless world volume modes. Cross section for this
process
yields agreement in scaling with the classical absorption in the $l$-th
partial wave. 

\newsec{Classical Absorption by Extremal Branes}

In this section we carry out classical absorption calculations for the
three cases of interest: the 3-brane in $D=10$, and the 2- and
5-branes in $D=11$.

The extremal 3-brane metric \hs\ can be written as  
$$ ds^2= A^{-1/2} \left (- dt^2 +dx_1^2+ dx_2^2+ dx_3^2\right )
+ A^{1/2} \left ( dr^2 + r^2 d\Omega_5^2 \right )
$$
where
$$A= 1+{R^4\over r^4}\ . 
$$
The s-wave of a minimally coupled massless scalar satisfies
\eqn\Coul{ \left [\rho^{-5} {d\over d\rho} \rho^5 {d\over d\rho} +
1 + {(\omega R)^4\over \rho^4} \right ] \phi(\rho) =0\ ,
}
where $\rho = \omega r$. Thus, we are interested in absorption by
the Coulomb potential in 6 spatial dimensions. 
For small $\omega R$ this
problem may be solved by matching an approximate
solution in the inner region to an approximate solution
in the outer region.

To approximate in the inner region, it is convenient to
use the variable $z=(\omega R)^2/\rho$. Then \Coul\ turns into
\eqn\Coulone{ \left [{d^2\over d z^2} - {3\over z} {d\over dz}
+1 + {(\omega R)^4\over z^4} \right ] \phi =0\ ,
}
Substituting $\phi = z^{3/2} f(z)$, we find
\eqn\Coultwo{ \left [{d^2\over d z^2} - {15\over 4 z^2} 
+1 + {(\omega R)^4\over z^4} \right ] f =0\ .
}
The last term may be ignored if $z\gg (\omega R)^2$, i.e. if
$\rho \ll 1$. In this region, \Coultwo\ is easily solved in terms of
cylinder functions. Since we are interested in the incoming wave for
small $\rho$, the appropriate solution is
\eqn\inner{\phi = i (\omega R)^4 \rho^{-2} 
\left [J_2 \left ({(\omega R)^2\over \rho}\right )+
i N_2 \left ({(\omega R)^2\over \rho}\right )  \right ] \ , 
\qquad \rho \ll 1\ ,
}
where $J$ and $N$ are the Bessel and Neumann functions.

Another way to manipulate \Coul\ is by substituting
$\phi = \rho^{-5/2} \psi$, which gives
\eqn\Coulthree{ [{d^2\over d \rho^2} - {15\over 4 \rho^2} 
+1 + {(\omega R)^4\over \rho^4}] \psi =0\ .
}
Now the last term is negligible for $\rho \gg (\omega R)^2$,
where \Coulthree\ is solvable in terms of cylinder functions.
If $\omega R\ll 1$, then the inner region ($\rho\ll 1$) overlaps
the outer region ($\rho \gg (\omega R)^2$), and the 
approximate solutions may be matched. We find that \inner\
matches onto
\eqn\outer{\phi = {32\over \pi}
\rho^{-2} J_2(\rho) \ , \qquad \rho \gg (\omega R)^2\ .
}
The absorption probability may be calculated as the
ratio of the flux at the throat to the incoming flux at
infinity, with the result
$$ {\cal P}= {\pi^2\over 16^2} (\omega R)^8 \ .$$
In $d$ spatial dimensions, the absorption cross-section is
related to the s-wave absorption probability by \dgm
$$ \sigma = {(2 \pi)^{d-1}\over \omega^{d-1} \Omega_{d-1}} {\cal P}
\ ,$$
where
$$\Omega_D = {2 \pi^{{D+1\over 2}}\over \Gamma \left (
{D+1\over 2}\right ) }
$$ 
is the volume of a unit $D$-dimensional sphere.
Thus, for the 3-brane we find\foot{By absorption cross section
we will consistently mean
the cross section per unit longitudinal volume of the
brane.}
\eqn\three{ \sigma_{\rm 3-brane}= {\pi^4\over 8}\omega^3 R^8 \ .} 

This exercise may be easily repeated for the other two
non-dilatonic branes.
For the M5-brane the extremal metric is \guv
$$ ds^2= A^{-1/3} \left (-dt^2 +dx_1^2+\ldots +  dx_5^2\right )
+ A^{2/3} \left ( dr^2 + r^2 d\Omega_4^2 \right )
$$
where $A = 1+ {R^3\over r^3}$.
Now the s-wave problem reduces to
absorption by the Coulomb potential in 5 spatial dimensions,
$$ \left [\rho^{-4} {d\over d\rho} \rho^4 {d\over d\rho} +
1 + {(\omega R)^3\over \rho^3}\right ] \phi(\rho) =0\ .
$$
The approximate solution in the inner region is
$$ \phi = i y^3 [J_3 (y) + i N_3 (y) ]
$$
where $y= 2 (\omega R)^{3/2}/\sqrt \rho$. This matches onto
$$\phi = 24 \sqrt {2\over \pi} \rho^{-3/2} J_{3/2} (\rho)
$$
in the outer region. The absorption probability is
${\cal P}=\pi (\omega R)^9/9$, and the absorption cross section is
found to be
\eqn\five{ \sigma_{\rm 5-brane}= {2 \pi^3\over 3}\omega^5 R^9 \ .} 
 
For the M2-brane the extremal metric is \ds
$$ ds^2= A^{-2/3} \left (-dt^2 +dx_1^2+ dx_2^2\right )
+ A^{1/3} \left ( dr^2 + r^2 d\Omega_7^2 \right )
$$
where $A = 1+ {R^6\over r^6}$. Now the s-wave problem reduces to
absorption by the Coulomb potential in 8 spatial dimensions,
$$ \left [\rho^{-7} {d\over d\rho} \rho^7 {d\over d\rho} +
1 + {(\omega R)^6\over \rho^6}\right ] \phi(\rho) =0\ ,
$$
Now the solution in the inner region is
$$ \phi = i y^{3/2} [J_{3/2} (y) + i N_{3/2} (y) ]
$$
where $y= (\omega R)^3/\rho^2$. This matches onto
$$\phi = 48 \sqrt {2\over \pi} \rho^{-3} J_3 (\rho)
$$
in the outer region. The absorption probability is
${\cal P}=\pi (\omega R)^9/24^2$, and the absorption cross section is
found to be
\eqn\two{ \sigma_{\rm 2-brane}= {2 \pi^4\over 3}\omega^2 R^9 \ .}

\newsec{Absorption of Scalars in the Effective Field Theory }

In this section we perform effective field theory calculations 
for the D3-branes and 
find complete agreement with the classical results.
For the M2-branes and
the M5-branes all the scaling exponents agree, but the normalizations cannot
be fixed due to insufficient knowledge of the world volume theory
describing many parallel branes.

First we consider absorption of scalars by D3-branes.
There are several types of fields that act as scalars from the point of 
view of the $D=7$ black hole obtained by wrapping the 3-brane over
$T^3$. For example, we could consider the gravitons with polarizations
along the 3-brane, $h_{\alpha\beta}$. Another scalar is the dilaton, 
$\phi$, which we discuss in detail here. 
From the low energy effective action of type IIB theory it is clear that
the dilaton is a minimally coupled massless scalar, i.e. its
s-wave part satisfies the
equation \Coul\ analyzed in the previous section. Our effective action
analysis will produce the absorption cross section which is in
perfect agreement with the classical result, \three, obtained from an
analysis of \Coul.

The coupling of the dilaton to the quadratic terms in the D3-brane
action is\foot{I am grateful to S. Gubser and A. Tseytlin for
valuable discussions on the structure of this action.} \Ark
\eqn\action{S= T_3
\int d^4 x\  
\left [{1\over 2} \sum_{i=4}^9 \partial_{\alpha} X^i
\partial^{\alpha} X^i - {1\over 4} e^{-\phi} F_{\alpha\beta}^2 \right ]
\ .}
where $F_{\alpha\beta}$ is the field strength for the gauge field
on the 3-brane describing its longitudinal dynamics, while
the 6 fields $X^i$ describe its transverse oscillations.
\eqn\threetension{
T_3=\sqrt \pi/\kappa_{10}
}
is the D3-brane tension \JP. A string theoretic calculation of all the
cubic terms in \action\ can be given with the methods developed in
\HK.

Fixing the gauge for $A_\alpha$, we find two physical photons.
Thus, there are 2 canonically normalized physical
fields $\tilde A$, each having a cubic 
coupling to the dilaton given by
$$ -{1\over 2 } \int d^4 x\ \phi 
\partial_{\alpha} \tilde A \partial^{\alpha} \tilde A
\ .$$
The 10-dimensional effective action is given by
$$ S_{bulk}= {1\over 2\kappa_{10}^2}
\int d^{10} x \sqrt g [R - {1\over 2}
\partial_\mu \phi \partial^\mu \phi + ...]
$$
so that the canonically normalized dilaton field is
$$ \tilde \phi = {\phi\over \sqrt 2 \kappa_{10}}
\ .$$
Thus, the world volume theory contains the coupling
\eqn\vertex{ -{\kappa_{10}\over \sqrt 2} 
\int d^4 x \  \tilde \phi 
\partial_{\alpha} \tilde A \partial^{\alpha} \tilde A
\ .}
A scalar incident on the brane at right angles 
may be converted into a pair of bosonic massless world volume modes moving
in opposite directions (it is easy to see that a pair of on-shell
fermions cannot be created).
Calculating the amplitude for this process, we get
$$ {\cal A} = -{\kappa_{10}\over \sqrt 2} 
2 {p_1\cdot p_2\over \sqrt 2 \omega^{3/2}}
=-{\kappa_{10}\sqrt\omega\over 2}\ .$$
Note that each state has normalization factor 
$1/\sqrt {2 E}$ and $E_1= E_2= \omega/2$. There is also a factor of 2
because either of the $\tilde X$'s can create either of the final particles.
Since $\vec p_1 = - \vec p_2$, 
$p_1\cdot p_2= \omega^2/2$.

Thus, each species contributes the absorption cross-section
$$ {1\over 2}\int {d^3 p_1\over (2\pi)^3} \int {d^3 p_2\over (2\pi)^3}
(2\pi)^4 \delta (E_1+E_2-\omega) 
\delta^3 (\vec p_1+\vec p_2) {\cal A}^2
$$
where ${1\over 2}$ is included because the final particles are identical.
Doing the integral we find that the absorption cross section
due to each species of massless bosons coupling to the dilaton is
\eqn\share{
{1\over 2} {\kappa_{10}^2 \omega^3 \over 32 \pi}
}
Since there are 2 such species, the total absorption cross-section
is
$$\sigma = {\kappa_{10}^2 \omega^3 \over 32 \pi}
\ .$$
So far, our analysis has covered the case of a single D3-brane.
Extending it to $N$ coincident D3-branes is straightforward.
Now each of the $\tilde A$'s is replaced by a hermitian 
$N\times N$ matrix, and the interaction vertex becomes\foot{
Terms involving $[\tilde A_\alpha, \tilde A_\beta]$ 
give contributions to the
cross section which are suppressed by powers of $\kappa_{10}$. Perhaps
they define the quantum corrections to the classical result of General
Relativity.}
$$ -{\kappa_{10}\over \sqrt 2} 
\int d^4 x\ \tilde \phi 
\Tr \  \partial_{\alpha}\tilde A \partial^{\alpha} \tilde A
\ .$$
Now there are $2 N^2$ possible species in the final state,
each contributing \share\ to the absorption cross section.
Thus, the string theoretic result for the total
cross section is
\eqn\EFT{\sigma_{\rm 3-brane} = {\kappa_{10}^2 N^2\omega^3 \over 32 \pi}
\ .}
Now we show that this is identical to the
classical result, \three.
We equate the ADM mass per unit volume of the 3-brane, 
$${2\pi^3 R^4\over \kappa_{10}^2}
\ ,$$
to the corresponding quantity in the D-brane description \JP,
$ {\sqrt \pi \over \kappa_{10} } N $.
Thus, we find \GKP
$$ R^4= {\kappa_{10} N\over 2\pi^{5/2}}
\ .$$
With this substitution, the classical formula, \three,
becomes identical to the string
theory result, \EFT. This is the first example 
outside the domain of validity of the effective string model
where such matching works exactly! 

Now we turn to the M-branes. Here we are forced to be more schematic
because a world volume description of multiple parallel branes is not
yet understood. We will simply assume a minimal coupling between
a scalar field and the massless world volume modes,\foot{
To estimate the scaling of the 
absorption cross section it is sufficient to
leave out the world volume gauge fields and to work with the scalars
only.}
\eqn\maction{ S= {T_p\over 2} \int d^{p+1} x\ \phi 
\partial_\alpha X^i \partial^\alpha X^i
\ . }
We also assume that the number of such modes scales with $N$ in the
way suggested by the near-extremal entropy, i.e. as $N^3$ for $N$
coincident M5-branes, and as $N^{3/2}$ for $N$
coincident M2-branes \kt. Introducing properly normalized fields,
$$ \tilde \phi \sim {\phi\over \kappa_{11}}\ , \qquad
\tilde X^i = \sqrt{T_p} X^i \ ,$$
we find the cubic vertex
$$S_3\sim  \kappa_{11} 
\int d^{p+1} x \  \tilde \phi 
\partial_\alpha\tilde X^i \partial^\alpha \tilde X^i
\ .$$
Calculating the cross-sections, we then have
\eqn\effive{
\sigma_{\rm 5-brane}\sim \kappa_{11}^2 \omega^5 N^3 \ ,}
\eqn\eftwo{
\sigma_{\rm 2-brane}\sim \kappa_{11}^2 \omega^2 N^{3/2} \ .}
In order to compare them with the classical results, we need
the charge quantization rules. For $N$ coincident M5-branes, we
have \refs{\KT,\kt}
$$ q_5 = N\sqrt 2 \left ({\pi\over 2\kappa_{11} }\right )^{1/3}
= {3\Omega_4\over \sqrt 2 \kappa_{11} } R^3\ .
$$
Solving for $R$ and substituting into the classical result, \five,
reduces it to \effive, up to normalization. For $N$ coincident M2-branes
\refs{\KT,\kt},
$$ q_2 = N\sqrt 2 (2 \pi^2\kappa_{11} )^{1/3}
= {6\Omega_7\over \sqrt 2 \kappa_{11} } R^6\ .
$$
Solving for $R$ and substituting into the classical result, \two,
reduces it to \eftwo, up to normalization.

\newsec{Higher Partial Waves }

In this section we estimate the classical cross-sections 
for scalars incident in higher partial waves. This allows us to
identify the terms in the world volume effective actions that
are responsible for these processes.

For the extremal 3-brane, the $l$-th partial wave satisfies
$$\left [\rho^{-5} {d\over d\rho} \rho^5 {d\over d\rho} +
1 + {(\omega R)^4\over \rho^4}- 
{l(l+4)\over \rho^2}\right ] \phi^{(l)} =0
$$
In the outer region, the approximate solution is
$$ \phi^{(l)} = B \rho^{-2} J_{l+2} (\rho)
\ ,$$
while in the inner region
$$\phi^{(l)} = i (\omega R)^4 \rho^{-2} 
\left [J_{l+2} \left ({(\omega R)^2\over \rho}\right )+
i N_{l+2} \left ({(\omega R)^2\over \rho}\right )  \right ] \ . 
$$
Matching the two regions, we find that 
$B \sim (\omega R)^{-2 l}$.
Therefore, the ratio of fluxes is
$\sim (\omega R)^{8+4l}$, and the absorption cross-section is
\eqn\higherthree{ \sigma^{(l)}_{\rm 3-brane}\sim \omega^{3+ 4l} R^{8+ 4l}\sim
\omega^{3+ 4l} (N \kappa)^{2+l} \ . 
}

Analysis of the effective action shows that
all partial waves are reproduced 
(at least schematically) by the leading term
in the effective action
$$ {\sqrt \pi\over 4 \kappa_{10}} \int d^4 x\ \phi (x, X) 
F^2_{\alpha\beta} \ .
$$
The term responsible for absorbing the $l$-th partial wave is\foot{
To obtain the 
correct normalization of the cross section, it is probably necessary
to add the fermionic terms required by supersymmetry.
In this paper we restrict ourselves to analyzing the purely bosonic 
processes.}
\eqn\abelian{ {\sqrt\pi \over 4  \kappa_{10}} \int d^4 x 
\ {1\over l!} (\partial_{i_1}\ldots  \partial_{i_l}\phi )
X^{i_1}\ldots X^{i_l} F^2_{\alpha\beta} 
\ .}
For $N$ coincident 3-branes, the natural non-abelian generalization 
of \abelian\ is
$$ {\sqrt\pi \over 4  \kappa_{10}} \int d^4 x 
\ {1\over l!} (\partial_{i_1}\ldots  \partial_{i_l}\phi )
\Tr \ X^{i_1}\ldots X^{i_l} F^2_{\alpha\beta} 
\ .  $$
It is not hard to see that the amplitude produced by this term 
scales as $\sim \kappa_{10}^{(2+l)/2}$ so that
the scaling of the cross section agrees with that of the
the classical result, \higherthree.
The number of distinct final states grows as
$N^{2+l}$, so that the $N$-dependence also agrees with
\higherthree.\foot{For example, the $U(N)$ index structure of 
the cubic ($l=1$) vertex is 
$ X^I_J \partial_\alpha X^J_K \partial^\alpha X^K_I
$. There are three independent summations giving the group theory factor
$\sim N^3$.}
Finally, a quick estimate of the $\omega$-dependence
from the Feynman rules gives $\omega^{3+4l}$, again in agreement
with \higherthree. Of course, once the power of $\kappa_{10}$ is matched, the
power of $\omega$ is guaranteed to be correct by dimensional analysis.

Now we extend this schematic analysis to the M-branes.
For the M5-brane the $l$-th partial wave satisfies
$$ \left [\rho^{-4} {d\over d\rho} \rho^4 {d\over d\rho} +
1 + {(\omega R)^3\over \rho^3}- {l(l+3)\over \rho^2}\right ] 
\phi^{(l)}(\rho) =0\ .
$$
In the inner region the approximate solution is
$$ \phi^{(l)} = i y^3 [J_{3+2l} (y) + i N_{3+2l} (y) ]
\ ,\qquad y= 2 (\omega R)^{3/2}/\sqrt \rho \ ,$$
which matches onto
$$\phi^{(l)} = B \rho^{-3/2} J_{(3+2l)/2} (\rho)
$$
in the outer region. 
We find that $B\sim (\omega R)^{-3l}$, so that
\eqn\higherfive{ \sigma^{(l)}_{\rm 5-brane}= \omega^{5+ 6l} 
R^{9+ 6l}\sim \omega^{5+ 6l} (N \kappa_{11}^{2/3})^{3+2l} \ .} 
 
For the M2-brane the $l$-th partial wave satisfies
$$ \left [\rho^{-7} {d\over d\rho} \rho^7 {d\over d\rho} +
1 + {(\omega R)^6\over \rho^6}
- {l(l+6)\over \rho^2} \right ] \phi^{(l)}(\rho) =0\ .
$$
Now the solution in the inner region is
$$ \phi^{(l)} = i y^{3/2} [J_{(3+l)/2} (y) + i N_{(3+l)/2} (y) ]
\ ,\qquad y= (\omega R)^3/\rho^2 \ ,$$
which matches onto
$$\phi^{(l)} = B \rho^{-3} J_{3+l} (\rho)
$$
in the outer region. 
We find that $B\sim (\omega R)^{-3l/2}$, so that
\eqn\highertwo{ \sigma^{(l)}_{\rm 2-brane}= \omega^{2+ 3l} 
R^{9+ 3l}\sim \omega^{2+ 3l} (N \kappa_{11}^{4/3})^{(3+l)/2} \ .} 
 
The scalings of
\higherfive\ and \highertwo\
with respect to $\kappa_{11}$ are reproduced
by the action \maction: we Fourier expand the scalar field 
$\phi(X)$ and identify the term with the $l$-th derivative 
of $\phi$ as the one responsible for absorbing the $l$-th partial
wave. The scalings with respect to $N$ are harder to understand.
We believe that they will provide valuable clues on the 
symmetry structure of
the effective action describing $N$ coincident M-branes.

\newsec{Conclusions}

The self-dual 3-brane of type IIB theory is a nice laboratory for comparing
string theory with semiclassical gravity. The stringy description
of a macroscopic 3-brane is well understood in terms of a large number
$N$ of parallel Dirichlet branes \refs{\JP,\EW}. The only scale
present in the low-energy effective action is 
$\kappa_{10} \sim g_{\rm str}\alpha'^2 $, 
the 10-dimensional gravitational constant.  
Using perturbative string theory, we may expand various quantities
in powers of $N\kappa_{10}\omega^4$. The result may be compared
with a similar expansion
generated by the semiclassical methods of General Relativity,
which use the classical 3-brane geometry as the background.
In this paper we have carried out such a comparison for the absorption
cross section of minimally coupled massless scalars, and found 
{\it exact} agreement to leading order. 

We believe that there is a number of interesting extensions of our
calculations. Comparing normalizations for higher partial waves
is a feasible, if somewhat technical, exercise.
Another interesting extension is to incident particles of higher
spin, such as the gravitons. In the effective field theory, the 
leading coupling of
the graviton of a given polarization, say $h_{67}$, is given by
$$\sqrt 2 \kappa_{10} \int d^4 x h_{67} 
\Tr \ \partial_{\alpha} \tilde X^6 \partial^{\alpha} \tilde X^7
$$
where $h_{67}$ is a canonically normalized field. 
The calculation is almost identical to that given in section 3,
and we find that the graviton is absorbed with the same 
cross section as the scalars, \EFT. In classical gravity, however, it
is quite difficult to derive the graviton propagation 
equations. It 
would be interesting to derive this equation in the 3-brane background
and compare the resulting absorption cross section 
with the prediction of string theory, \EFT.

The 2- and 5-branes of M-theory bear many similarities with the
3-brane of type IIB \kt. Their geometries are non-singular, while
their world volume theories are governed by the 
11-dimensional Planck scale. Thus, it should be possible to
compare the expansions in powers of $\kappa_{11}$ generated  
by the M-theory and the semiclassical supergravity.
We showed that, with minimal assumptions about
the world volume theories of many coincident M-branes, the
scalar absorption cross sections agree up to normalizations. 
If we assume that the exact agreement
must hold, then the information provided by the semiclassical methods
is a valuable guide to formulating the M-theory.

If the multiple coincident D3-branes and M-branes are indeed the
unitary quantum systems underlying their classical geometry, then
there is a wealth of perturbative calculations, of
the type carried out in \refs{\us,\gm,\HK} that may shed more
light on this remarkable phenomenon.

\newsec{Acknowledgements}

We are grateful to C.G.~Callan, S.S.~Gubser, M.~Krasnitz and
A.A.~Tseytlin for useful discussions and comments.  This work 
was supported in part by DOE grant DE-FG02-91ER40671,
the NSF Presidential Young Investigator Award PHY-9157482, and the
James S.{} McDonnell Foundation grant No.{} 91-48. 

\listrefs
\bye